\documentstyle[prd,aps,preprint,epsfig]{revtex}

\tighten

\begin{document}
\draft
 
\pagestyle{empty}

\preprint{
\noindent
\hfill
\begin{minipage}[t]{3in}
\begin{flushright}
LBNL--43001 \\
UCB--PTH--99/10\\
March 1999
\end{flushright}
\end{minipage}
}

\title{Final State Interaction Phase in B Decays}

\author{
Mahiko Suzuki
}
\address{
Department of Physics and Lawrence Berkeley National Laboratory\\
University of California, Berkeley, California 94720\\}
\author{
Lincoln Wolfenstein}
\address{
Department of Physics, Carnegie Mellon University, Pittsburgh, PA 15213
}


\maketitle

\begin{abstract}

From an estimate of the meson-meson inelastic scattering at 5 GeV
it is concluded that a typical strong phase in B decays to two mesons
is of order of $20^{\circ}$.  For a particular final state an estimate 
of the phase depends on whether that state is more or less probable 
as a final state compared to those states to which it is connected by
the strong interaction S matrix.

\end{abstract}
\pagestyle{plain}
\narrowtext

\setcounter{footnote}{0}

\section{Introduction} 

Strong final state interactions play an important role in the analysis of 
CP-violating effects in B decays.  Direct CP violation such as a difference 
in rates for $B^+\rightarrow F$ and $B^-\rightarrow \overline{F}$ vanishes
in the limit that there are no strong phase shifts.  Final state phases 
play a critical role in amplitude analyses of a set of $B^0$ decay 
experimental results.

An approach to final state phases in inclusive decays was given by Bander, 
Silverman, and Soni\cite{BSS}.  For decays corresponding to the transition
$b\rightarrow u\overline{u}s$ they considered at the quark level $b
\rightarrow c\overline{c}s \rightarrow u\overline{u}s$ where the second 
transition on mass shell yielded the final state phase.  Whether or not 
this is reasonable for inclusive decays, its application \cite{GH} to 
exclusive decays such as $B\rightarrow \overline{K}\pi$ has been 
criticized\cite{W} because the major final state interactions of 
$\overline{K}\pi$ are ``soft'' scattering to $\overline{K}n\pi$ and not to 
$c\overline{c}$ states. 

We concentrate here on the decay of $B$ to two mesons, referring to 
$B\rightarrow\pi\pi$ to be specific.  Arguments have been given that the 
final state phase shifts should be small. For example, Bjorken\cite{BJ} 
argues that there is little final state scattering in $B\rightarrow\pi\pi$
because the $B$ decays directly to colorless $q\overline{q}$ pairs that 
do not interact as they evolve into $\pi\pi$. Taken literally this is not 
correct since the $s$-state scattering at 5 GeV is expected to be 
sizable\cite{Dono} 
as we discuss below. In the present note we seek to analyze the relations 
between the weak decay amplitude and the strong interaction S-matrix of
final states that might lead to large or small final state phases.

\section{Multichannel final state interaction}

   Consider the decay matrix element 
$\langle f^{out}|\mbox{$\cal O$}_i|B\rangle$ for the 
$B$ meson into the hadronic final state $f$, where $\mbox{$\cal O$}_i$ is
a weak decay operator.  The strong interaction 
S-matrix is defined with the ``in'' and ``out'' states by
\begin{equation}
        S_{ff'} = \langle f^{out}|f'^{in}\rangle.  \label{S}
\end{equation} 
We choose states that are eigenstates of $J$ not of individual meson momenta.
The phases of the {\it in} and {\it out} states are fixed by the time 
reversal transformation T:
\begin{eqnarray}
              T|f'^{in}\rangle  &\rightarrow&  \langle f'^{out}|, \\ \nonumber
              T|f^{out}\rangle  &\rightarrow&  \langle f^{in}|.\label{T}
\end{eqnarray}
With this phase convention, time reversal invariance of strong interactions 
requires that $S_{ff'}$ be symmetric matrix:
\begin{equation}
               S_{ff'} = S_{f'f}. \label{symmetric}
\end{equation}  
 Applying time-reversal operation on 
$M_f=\langle f^{out}|\mbox{$\cal O$}_i|B\rangle$, one obtains
\begin{equation}
 M_f \stackrel{T}{\rightarrow}\langle B|T\mbox{$\cal O$}_iT^{-1}|f^{in}\rangle.
\end{equation}
If one inserts a complete set of {\it out} states and uses 
Eq.(\ref{symmetric}),
this relation becomes $M_f = \sum_{f'} S_{ff'}M_{f'}^*$ for a T-even
decay operator $\mbox{$\cal O$}_i$. One can express it in the operator form as
\begin{equation}
        M = SM^*,  \label{time-reversal}
\end{equation}
where $M$ is represented in a column vector.
This matrix equation is formally solvable as
\begin{equation}
       M = S^{1/2}M^0, \label{solution}
\end{equation}
where $M^0$ is an arbitrary {\it real} vector of the same dimension as $M$.
If one uses the eigenstates $|\alpha\rangle$ of the S-matrix as a basis,
Eq.(\ref{solution}) reduces to the Watson theorem: $M_{\alpha} = 
M_{\alpha}^0e^{i\delta_{\alpha}}$.  We thus may consider the vector $M^0$ as
representing the decay amplitude in the absence of the final state phases
due to the strong interaction.
Since $M$ and $M_0$ are related by a unitary matrix, it holds that
$\sum_f |M_f|^2 = \sum_f|M_f^0|^2$.

If one subtracts the complex conjugate of $M$ from both sides in 
Eq.(\ref{time-reversal}) and divides by $2i$, the familiar form 
${\rm Im}M = {\bf t}M^*$  emerges for the imaginary part of $M$, where
${\bf t} = (S-1)/2i$. In components, it reads
\begin{equation}
   {\rm Im}M_f = \sum_{f'}t_{ff'}M_{f'}^*.  \label{reduction} 
\end{equation}
This form is commonly derived starting with Lehmann-Symanzik-Zimmermann's
reduction formula. In applications of interest, the weak decay Hamiltonian 
$H_w$ is given in the form
\begin{equation}
     H_w = \sum_i\lambda_i\mbox{$\cal O$}_i, \label{operator}
\end{equation}
where $\lambda_i$ is a combination of the CKM matrix elements and 
$\mbox{$\cal O$}_i$ is a T-even operator. It is to be 
understood that $M_f$ is to be evaluated separately for different 
operators $\mbox{$\cal O$}_i$. 

\section{Strong interaction S-matrix}
  When two mesons such as $\pi^+\pi^-$ interact in the $s$-state, we 
believe that they will scatter into a large number of multi-particle final 
states.  Indeed we expect similar inelastic behavior for all partial waves of 
$l < kr$ where $r$ is a characteristic hadron radius.  The sum over these 
partial waves can be described by a diffractive scattering formula
such as that given by Pomeron exchange. For the case of meson-meson 
scattering we extrapolate from the analysis of meson-baryon and 
baryon-baryon scattering and write for the invariant elastic scattering
amplitude ,
\begin{equation}
          T(s,t) = i\sigma_{tot}se^{bt}, \label{Pomeron}
\end{equation}   
where the constant in front has been fixed by the optical theorem.
Defining the Pomeron residue by $\beta(t)\equiv \sigma_{tot}e^{bt}$, 
we obtain with the factorization relation $\beta(t)_{MM'}\beta(t)_{pp}=
\beta(t)_{Mp}\beta(t)_{M'p}$ 
\footnote{The factorization can be proved only for simple $l$-plane 
singularities. It is an assumption for the Pomeron.}
\begin{equation}
    \sigma_{tot}^{\pi\pi} \simeq 12{\rm mb},\;\;\; 
    \sigma_{tot}^{\pi K}\simeq 10{\rm mb},\;\;\;
    \sigma_{tot}^{K\overline{K}}\simeq 8{\rm mb},
\end{equation}
where $\sigma_{tot}^{pp}= 37$mb, $\sigma_{tot}^{\pi p}=21$mb, and 
$\sigma_{tot}^{Kp}= 17$mb\cite{Barger}
have been used for the diffractive contribution 
of $\sigma_{tot}$ at $\sqrt{s} \simeq m_B$.  For the diffractive peak width, 
the factorization gives
\begin{equation}
 b^{\pi\pi} \simeq 3.6{\rm GeV}^{-2}, \;\;\;
 b^{K\pi} \simeq 2.8{\rm GeV}^{-2},\;\;\;
 b^{K\overline{K}} \simeq 2.0{\rm GeV}^{-2}
\end{equation}
if we choose $b^{pp}\simeq 5.0{\rm GeV}^{-2}$, $b^{\pi p}\simeq 
4.3{\rm GeV}^{-2}$,
and $b^{Kp}\simeq 3.5{\rm GeV}^{-2}$\cite{CR}.  For $D\pi$ scattering, we 
use the quark counting rule for $\sigma_{tot}$ and the assumption 
that the charmed quark interacts with the light quarks much more weakly.
Then we obtain a crude estimate
\begin{equation}
  \sigma_{tot}^{D\pi}\approx \frac{1}{2}\sigma_{tot}^{\pi\pi},
\end{equation}
and $b^{D\pi}$ is a little smaller than $b^{K\pi}$. 

 Projecting out the $s$-wave from the  
amplitude in Eq.(\ref{Pomeron}),
\begin{equation}
    a_{l=0}(s) = \frac{1}{16\pi s}\int_{-s}^0 T(s,t)dt \label{Projection}
\end{equation}
yields
\begin{equation}
  {\rm Im}a_{l=0}\simeq \left\{\begin{array}{cc}
                 0.16 &(\pi\pi)\\
                 0.17 &(K\pi)\\
                 0.18 &(K\overline{K})\\
                 0.12 &(D\pi) \end{array} \right. \label{15}
\end{equation}
at $\sqrt{s} = 5\sim 6 $GeV.
 
Extraction of the $s$-wave amplitude from the diffractive formula 
may arouse suspicion since one thinks of diffraction as a peripheral 
process.\cite{GR} It would be better to consider $T(s,t)$ as 
describing the scattering from an absorbing gray sphere of radius $r$.
The values of $a_l$ up to  $l \sim kr$ vary gradually with $l$ thus adding 
up to a large forward peak.  As a result about 90\% of the contribution 
to the integral in Eq.(\ref{Projection}) comes from $|t|< 1 {\rm GeV}^2$
even though the $l=0$ amplitude itself is, of course, independent of $t$.

In what follows we use the estimate 
\begin{equation}
        S_{ff} \simeq 0.7  \label{Sff}
\end{equation}
corresponding to
\begin{equation}
    a_{l=0} \equiv t_{ff} = \frac{S_{ff}-1}{2i}=0.15 i.\label{New}
\end{equation}
This corresponds to the case of a gray sphere with an inelasticity of
0.85. In the limiting case of a black sphere $S_{ff}$ goes to zero and
the inelasticity goes to 0.5. If one goes beyond
the diffractive scattering approximation, $S_{ff}$ is not purely real. In the 
Regge theory, the real part arises from exchange of the non-Pomeron 
trajectories such as $\rho$ and $f_2$. In $\pi^{\pm}p$ scattering the 
real-to-imaginary ratio of $10\sim 20\%$ was observed in the forward
scattering amplitudes at $\sqrt{s}= 5\sim 6$ GeV.\cite{Foley} 
We can make an estimate of the real part for meson-meson scattering by
using the factorization. We first determine the Regge parameters at $t=0$ 
from the total cross section differences\cite{Barger} and then extract their 
$t$-dependences from the angular dependence of the differential 
cross sections.\cite{BO} The analysis is simpler if exchange degeneracy
is imposed.  The smaller $\sigma_{tot}$ and the larger $\rho$-$f_2$ 
residues tend to enhance the real to imaginary amplitude ratio for $\pi\pi$ 
scattering over $\pi p$ scattering, while the smaller $b^{\pi\pi}$ partially 
compensates the trend.  Particularly for $\pi^+\pi^0$, the real parts of the
$\rho$ and $f_2$ terms add up close to $30\%$ of the imaginary part.
However, our major goal is to understand the implications of the sizable
inelastic scattering; for this purpose we use the simplifying approximation 
that $S_{ff}$ is real.

\section{Two channels}
  The relation (\ref{time-reversal}) 
was studied in the case of two channels \cite{Dono}
assuming that the diagonal S-matrix elements $S_{ff}$ 
are purely real. 
This requirement on the S-matrix turns out to be so strong
in the case of two channels that there is only a single parameter left:
\begin{eqnarray}
   S&=& \left( \begin{array}{cc}    \frac{1}{\sqrt{2}}& -\frac{1}{\sqrt{2}}\\
                                  \frac{1}{\sqrt{2}}& \frac{1}{\sqrt{2}}
                     \end{array}\right)\left( \begin{array}{cc}
                                  e^{2i\theta}& 0 \\
                                  0 & e^{-2i\theta}\\
                     \end{array}\right)\left( \begin{array}{cc}
                                  \frac{1}{\sqrt{2}}& \frac{1}{\sqrt{2}}\\ 
                                -\frac{1}{\sqrt{2}}& \frac{1}{\sqrt{2}}
                     \end{array}\right), \\ \nonumber
    &=& \left( \begin{array}{cc} \cos 2\theta & i\sin 2\theta \\
                                 i\sin 2\theta & \cos 2\theta \\
                     \end{array} \right).     \label{twochannel}
\end{eqnarray}            
When $S^{1/2}$ is computed from $S$ and substituted in Eq.(\ref{solution}),
a simple relation results:
\begin{eqnarray}
      M_1 &=& M_{01}\cos\theta +iM_{02}\sin\theta, \\ \nonumber 
      M_2 &=& iM_{01}\sin\theta +M_{02}\cos\theta.  \label{MMzero}
\end{eqnarray}  
The phase of the decay amplitude in the channel 1 is large if the particle
decays preferrentially to the channel 2, while it is small if the channel 1
is the dominant decay channel for small values of $\theta$. 
Though it is an interesting conclusion,
this picture turns out to be specific to the two-channel case.  When we
add one more channel in the final state, there are three real parameters
even after Im$S_{ff}=0$ is assumed. The nice simple
relation of Eq.(18) does not hold any longer. If we go to $N$
final channels, there are $N(N-1)/2$ real parameters even with 
Im$S_{ff}=0$, and no meaningful prediction results.
Therefore we must change our strategy in 
studying the case of $N \gg 1$ such as the $B$ decay. 

\section{Randomness of weak decay amplitudes}
  In the presence of many decay channels, strong interactions are so 
complicated that it is beyond our ability to predict final state 
interactions accurately. We must substitute lack of our knowledge 
with reasonable dynamical assumptions and/or approximations. In search of 
such an assumption, we notice that since $M_{f'}$ and $S_{ff'}$ come from
two different sources, weak and strong interactions, the phase of 
product $S_{f'f}M_{f'}^*$ for $f'\neq f$ takes equally likely a positive 
or a negative value as $f'$ is varied with $f$ fixed. While $M_{f'}$ is 
related to $M_{f}$ ($f'\neq f$) by rescattering, there exist so many states 
that the influence of $f$ on $f'$ can be disregarded.
We therefore introduce the postulate that the phase of $S_{ff'}M_{f'}^*$ 
takes random values as $f'$ varies.  
It should be noted that randomness is postulated here for the relative phase 
and sign of the decay matrix element to the S-matrix element, not for the 
dynamical phases and mixing of S-matrix as it was introduced 
in the random S-matrix theory of nuclear physics.\cite{Dyson}

  We start our analysis with Eqs.(\ref{reduction}) and (\ref{Sff}). Taking 
the $f'=f$ term in the sum to the left-hand side in Eq.(\ref{reduction})
and using $t_{ff}\simeq i{\rm Im}t_{ff}$, 
we write Eq.(\ref{reduction}) in the form
\begin{equation}
   (1+it_{ff}){\rm Im}M_f - t_{ff}{\rm Re}M_f = 
    \sum_{f'\neq f} t_{ff'}M_{f'}^*. \label{Formula}
\end{equation}
The first and the second term of the left-hand side are real and imaginary,
respectively, for Re$t_{ff} = 0$. Given the estimate of Eq.(\ref{New}) the
coefficient of the first term is much larger than that of the second term 
and we consider this primarily as an equation for Im$M_f$.
With the randomness postulate, 
the phase of $M_f$ is equally often positive or negative
if we consider a large ensemble of final states $f$. It is some kind of 
average of the magnitude of the phase of $M_f$, not values for individual
$M_f$, that we can study with our randomness postulate. 
For this purpose, we take the absolute
value squared for both sides of Eq.(\ref{Formula}).
Then the right-hand side is:
\begin{equation}
 {\rm RHS} = \frac{1}{4}\sum_{f',f''\neq f}S_{ff'}M_{f'}^*S_{ff''}^*M_{f''},
\end{equation}         
where $t_{ff'}= S_{ff'}/2i$ has been used for $f'\neq f$.
The random phase postulate allows us to retain only the terms of $f'=f''$ and
to reduce the double sum to a single sum:
\begin{eqnarray}
{\rm RHS} &\simeq&\frac{1}{4}\sum_{f'\neq f}S_{ff'}S^{\dagger}_{f'f}|M_{f'}^2|
              \\ \label{Inter}
  &\equiv& \frac{1}{4}\overline{|M_{f'}^2|}\sum_{f\neq f'}|S_{ff'}|^2,
         \label{Intermediate}
\end{eqnarray}
where the second line defines $\overline{|M_{f'}^2|}$ as the weighted average
of the decay amplitudes into states $f'$. Then, using the unitarity 
of S-matrix, we reach
\begin{equation}
   {\rm RHS} \simeq \frac{1}{4}(1 - S_{ff}^2)\overline{|M_{f'}^2|}.
      \label{Intermediate2}
\end{equation}
While our estimate of $S_{ff}$ is made on the basis of the Pomeron exchange, 
it should be noted that contributions to $S_{ff'}$ from quantum number 
exchange may be important in determining $\overline{|M_{f'}^2|}$ from
Eq.(\ref{Intermediate}) if they correspond to states $f'$ with large values
of $|M_{f'}^2|$. Identifying 
Eq.(\ref{Intermediate2}) with the absolute value squared of the left-hand 
side of Eq.(\ref{Formula}),
we obtain the prediction of our random phase approximation:
\begin{equation}
   (1+S_{ff})^2({\rm Im}M_f)^2 + (1-S_{ff})^2({\rm Re}M_f)^2 =
             (1- S_{ff}^2)\overline{|M_{f'}^2|}. \label{Prediction}
\end{equation}
Defining $\rho$ by
\begin{eqnarray}
   \rho &\equiv& \overline{|M_{f'}^2|}^{1/2}/|M_f|, \\ \nonumber
   |M_f|^2 &=& ({\rm Im}M_f)^2 +({\rm Re}M_f)^2,
\end{eqnarray}
the ratio of the imaginary-to-real part of $M_f$ 
is solved from Eq.(\ref{Prediction}) as
\begin{equation}
   \frac{({\rm Im}M_f)^2}{({\rm Re}M_f)^2} \equiv \tan^2\delta_f
    = \frac{\tau^2(\rho^2 - \tau^2)}{1-\rho^2\tau^2},  \label{Ratio}
\end{equation}
where
\begin{equation}
    \tau = \biggl(\frac{1-S_{ff}}{1+S_{ff}}\biggr)^{1/2}.
\end{equation}
Note that $\tau^2$ is equal to the ratio of elastic to inelastic scattering
cross section $\sigma_{el}/\sigma_{inel}$  of the relevant partial wave. 
   Since the left-hand side of Eq.(\ref{Ratio}) is nonnegative, $\tau$
and $\rho$ are constrained for $S_{ff}>0$ by
\begin{equation}
        \tau^2 \leq \rho^2 \leq 1/\tau^2. \label{Constraint}
\end{equation}
For $S_{ff}=0.7$, 
\begin{equation}
      \tau^2 = 0.18,
\end{equation}
so that rescattering among the final states does not allow
$\overline{|M_{f'}|^2}$ and $|M_f|^2$ to differ too greatly
in magnitude. In the weak limit of rescattering ($\tau\rightarrow 0$), 
of course, Eq.(\ref{Constraint}) allows any value for $\rho$.
In the black sphere limit ($\tau\rightarrow 1$) Eq.(\ref{Ratio}) is useless
and Eq.(\ref{Constraint}) constrains $\rho=1$.  Our approach is only useful 
to the extent that inelastic scattering dominates very much over elastic  
for the final state $f$. It should be noticed that Eq.(\ref{Ratio}) reduces to 
the two-channel case Eq.(18) with
\begin{equation}
  \biggl|\frac{M_2^0}{M_1^0}\biggr|^2 = \frac{\rho^2-\tau^2}{1-\rho^2\tau^2}.
\end{equation}
Our random approximation amounts to lumping all inelastic channels together as 
if they were a single inelastic channel with an ``average'' decay amplitude.
However, we now interpret this as something like the standard deviation of
the phase for an ensemble of independent final states $f$ with a given value
of $\rho$.

If the relevant states $f'$ were similar to the state $f$, 
then we might expect $\rho$ to be close to unity. For $\rho=1$ Eq.(\ref{Ratio})
reduces to
\begin{eqnarray}
          \tan^2\delta_f &=&\tau^2
            = \frac{1-S_{ff}}{1+S_{ff}},\\ \nonumber
          \sin\delta_f &=& \sqrt{\frac{1-S_{ff}}{2}}.  \label{delta}
\end{eqnarray}
With $S_{ff}= 0.7$ this gives $|\delta_f| \sim 23^{\circ}$.  Thus a
typical value of the final strong interaction phase in this case is not
small. This result for a typical state has a simple heuristic interpretation.
The original real decay amplitude $M_1^0$ is reduced as a result of 
absorption by a factor $a$, but an imaginary term arises due to rescattering 
from other states.  Since the total decay rate is not changed by final-state
scattering the final value of $|M_f|$ for a typical state will be equal to
$|M_1^0|$.  Thus $M_f$ takes the form
\begin{eqnarray}
     M_f &=& M_1^0[a + i\sqrt{1-a^2}], \\ \nonumber
     \frac{{\rm Im}M_f}{{\rm Re}M_f} &=& \sqrt{1-a^2}/a. \label{W1}
\end{eqnarray}
This agrees with the result above if the absorption factor is identified as
\begin{equation}
   a = \sqrt{(1+S_{ff})/2}. \label{W2}
\end{equation}

Any argument that a final state phase is small must be an argument that $\rho$
is small. It should be noted that $\rho$ depends on the particular final state
$f$ and on the weak interaction operator $\mbox{$\cal O$}_i$.  
The quantity $\overline{|M_{f'}^2|}$ 
is an average of the square of the decay amplitude to state $f'$ via 
$\mbox{$\cal O$}_i$ weighted by the square of the scattering amplitude from 
$f$ to $f'$ (cf Eqs.(\ref{Intermediate}) and (\ref{Intermediate2})).  
Thus a value of $\rho$ much smaller than unity
means that on average the states to which $f$ scatters are much less likely 
than $f$ to be final states in the decay due to operator $\mbox{$\cal O$}_i$.
Conversely, if $f$ is a particularly unfavored final state $\rho$ may well
be above unity.  Figure 1 shows the dependence of the phase on $\rho$ for
the choice of $S_{ff}=0.7$.

\section{Possibility of small or large strong phases for two-body channels}
While our randomness postulate leads to sizable strong phases for 
``typical'' decay channels, dynamical arguments are necessary to 
estimate the parameter $\rho$ for a specific channel and a specific
decay operator.  We ask what dynamical arguments could let 
our prediction agree with Bjorken's argument which favors  
small phases for decays such as $B^0\rightarrow \pi^+\pi^-$.
The observed branching fraction of $B^0\rightarrow\pi^+\pi^-$ 
is about $10^{-3}$ of the inclusive branching fraction for $b\rightarrow 
d\overline{q}q$ within large uncertainties.
This might seem to indicate that $\pi^+\pi^-$ is not a favored final 
state and that the strong phase of $\pi^+\pi^-$ might be of 
order of $20^{\circ}$. However this conclusion 
can be evaded if certain conditions are satisfied.

It should be noted that we are only interested in whether $\pi^+\pi^-$
is a favored decay channel relative to those to which it is connected by 
$S_{f'f}$. Most of the states $f'$ are multi-meson states with little jet-like 
character.  It could be argued that these states are not favored as $B$ 
decay products because they are not likely to develop from three quark jets 
into which $B$ naturally decays.   It is
not clear whether this distinction is really operative for the energy $m_B$.

 If the above argument were true, it
could be considered as an interpretation of the Bjorken argument.  In order
to produce a $\pi\pi$ final state the final quarks must emerge as colorless
pairs; alternative quark configurations are unlikely to hadronize  into
$\pi\pi$. Thus $\rho$ would be close to its minimum value and
Im$M_f$ would be small.  It is not true, however, that the
final $\pi\pi$ state has little interaction, but the effect would
primarily be a moderate absorption correction to the real part. In the 
two-channel reduction this corresponds to $M_{20}/M_{10}$ close to zero and 
the change of the real part of $M_1$ from $M_{10}$ to $M_{10}\cos\theta$
would be considered as the absorption correction.

An alternative possibility in which one might expect a large final state 
phase shift has been emphasized in a number of recent papers\cite{X}.  
These are cases in which $M_f$ vanishes in the naive factorization
approximation.  An example is the ``tree'' amplitude proportional to 
$\lambda_u = V_{ub}V_{us}^*$ for the decay $B^-\rightarrow\overline{K}^0\pi^-$.
  In this case the major contribution to
$M_f$ is expected to arise from  rescattering from favored states so that 
$\rho$ might be larger than unity. It can be argued\cite{X} that
in this case, even though the final state scattering mainly goes to 
multi-particle states, the main contribution to the strong phase arises
from the scattering of quantum number exchange
involving two particle to two particle transitions 

\section{Conclusion}

  For a ``typical'' final state the strong final-state phase shift
is not small; a typical magnitude is $20^{\circ}$.  We understand this to 
be the magnitude averaged over the states that are interconnected by the
final-state $S$-matrix.  We expect there to be sizable fluctuations about 
the average; in fact, since we cannot predict the sign of the phases, our 
analysis suggests that the algebraic average phase may be zero.

A simple heuristic understanding of this phase is that the final state 
absorption reduces the value of the original real decay amplitudes whereas 
rescattering from other states provides an imaginary amplitude.  For a 
``typical'' state the absolute value of the amplitude is not changed since
the final state interaction does not change the total decay rate.  Our 
magnitude estimate of $\sim 20^{\circ}$ is derived from the expected 
inelasticity of the meson-meson scattering.

   It may be possible to argue for a particular final state $f$ 
that the phase is small.  Any such argument must show that the states to 
which $f$ is connected by the $S$-matrix are generally less likely to be
a decay product of $B$ than $f$.  Thus it might be argued that  
the four-quark operator leads easily to $\pi^+\pi^-$ whereas there are 
many states to which $\pi^+\pi^-$ scatters that are not easily reached 
directly via $B$ decay.
Conversely it might be argued that for states which are not easily reached 
via the four-quark operator, the strong phase is large. 

\acknowledgements

 One of the authors (LW) acknowledges Miller Institute for 
Basic Research in Science for a Visiting Miller Research Professorship
during this work at Berkeley. He is also supported by the U.S. Department
of Energy under Contract No. DE-FG02-91-ER-40682.  The other author (MS) was 
supported in part by the Director, Office of Energy Research, Office of 
High Energy and Nuclear Physics, Division of High Energy Physics of the U.S.
Department of Energy under Contract No.DE-AC03-76SF00098 and in part
by the National Science Foundation under Grant No. PHY-95-14797.


\noindent
\begin{figure}
\epsfig{file=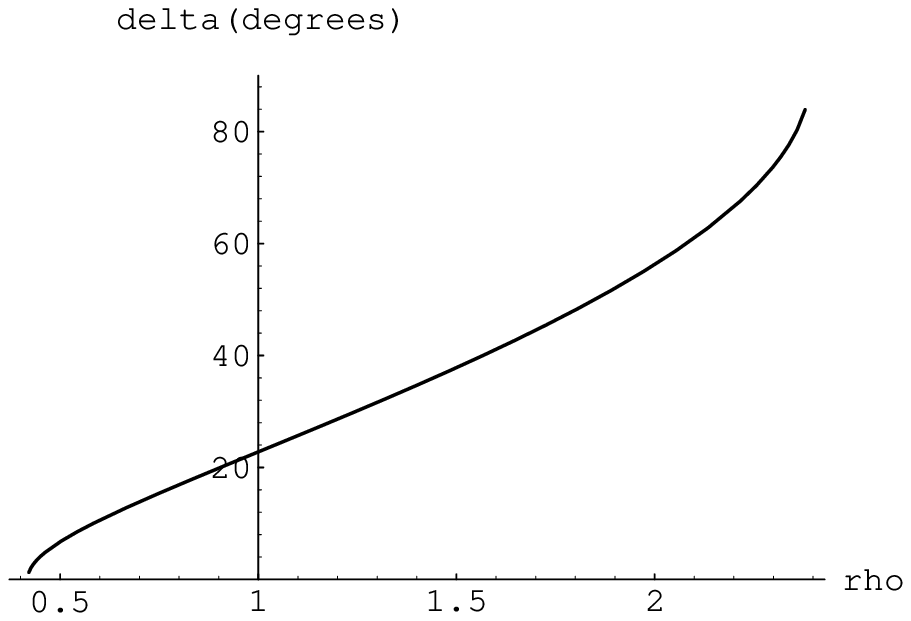,width=7cm,height=5cm}
\caption{The strong phase $\delta_f$ defined in Eq.(26) plotted against
the ratio $\rho$ for $S_{ff}=0.7$. $\delta_f$ is chosen between 
$0^{\circ}$ and $90^{\circ}$.
\label{fig:1}} 
\end{figure}
 
\end{document}